\documentclass[cameraready]{Interspeech}

\title{Enroll-on-Wakeup: A First Comparative Study of Target Speech Extraction for Seamless Interaction in Real Noisy Human-Machine Dialogue Scenarios}

\author[affiliation={1}]{Yiming}{Yang}
\author[affiliation={2}]{Guangyong}{Wang}
\author[affiliation={2}]{Haixin}{Guan}
\author[affiliation={1},correspondingauthor]{Yanhua}{Long}

\address{
    $^1$ Shanghai Normal University, Shanghai, China \\
    $^2$ Unisound AI Technology Co., Ltd., Beijing, China 
}

\email{y2379286479@outlook.com,  yanhua@shnu.edu.cn}

\keywords{Target speech extraction, seamless interaction, EoW-TSE}

\usepackage{comment}
\usepackage{amssymb}
\usepackage{graphicx}
\usepackage{CJKutf8}

\usepackage{cite}
\usepackage[skip=5pt]{caption}

\begin{document}

\maketitle

\begin{abstract}
Target speech extraction (TSE) typically relies on pre-recorded high-quality 
enrollment speech, which disrupts user experience and limits feasibility in 
spontaneous interaction. In this paper, we propose Enroll-on-Wakeup (EoW), 
a novel framework where the wake-word segment, captured naturally during 
human-machine interaction, is automatically utilized as the enrollment reference. 
This eliminates the need for pre-collected speech to enable a seamless experience. 
We perform the first systematic study of EoW-TSE, evaluating advanced 
discriminative and generative models under real diverse acoustic conditions. 
Given the short and noisy nature of wake-word segments, we investigate 
enrollment augmentation using LLM-based TTS. Results 
show that while current TSE models face performance degradation in EoW-TSE\footnote{https://github.com/Yym-line/EoW-TSE}, 
TTS-based assistance significantly enhances the listening experience, 
though gaps remain in speech recognition accuracy. 

\end{abstract}

\vspace{-0.3cm}
\section{Introduction}
\label{sec:intro}

Target speech extraction (TSE) aims to isolate a specific speaker's voice from 
a multi-talker or noisy acoustic environment by leveraging a set of auxiliary clues, 
such as a reference enrollment utterance from the target speaker. Traditional TSE 
frameworks typically operate under the assumption that a high-quality, 
pre-recorded enrollment signal is readily available to guide the extraction backbone 
in characterizing the speaker's unique embedding \cite{vzmolikova2019speakerbeam,Wang2018VoiceFilterTV,Wang2022WespeakerAR}. However, in practical human-machine 
dialogue scenarios, requiring users to provide a pre-collected enrollment sample 
beforehand significantly disrupts the fluidity of interaction and limits the 
system's feasibility for spontaneous or first-time users. To bridge this gap and 
enable truly seamless interaction, it is essential to move toward an ``Enroll-on-Wakeup'' (EoW) TSE.
In this setting, the system must extract the target speech using only the brief and often 
noise and interferer corrupted wake-word segment captured during the initial 
triggering phase, posing a significant yet necessary challenge for current TSE research.

Existing methods for providing target speaker clues in TSE can be broadly categorized 
into three paradigms: audio-only, audio-visual (AV-TSE), and spatial-assisted auditory TSE. 
Audio-only-enrollment based TSE primarily relies on pre-trained speaker verification models 
\cite{Desplanques_2020} or dedicated speaker encoders \cite{Xu_2020, Ge2020SpExAC, Chen2023MCSpExTE} 
to extract target speaker identity embeddings. Recently, speaker embedding-free  
methods \cite{11012711, huang2025sef, yang2024target} and waveform-level concatenation 
approaches \cite{Shen2025ListenTE} have demonstrated robust performance across various TSE tasks. Despite their success, these methods remain dependent on the 
requirement of pre-recorded enrollment. To leverage multi-modal information, 
AV-TSE \cite{Lin2023AvSepformerCS, 11209435, pan2023scenario} incorporates visual cues such as 
lip movements, while recent works \cite{wu2025elegance} integrate the linguistic knowledge 
of large language models to compensate for acoustic degradation. However, AV-TSE is 
fundamentally limited by line-of-sight requirements and raises significant privacy and 
computational concerns for low-power embedded devices. Alternatively, spatial-assisted 
TSE \cite{9746221, alcala2025location} utilizes multi-channel features or direction-of-arrival 
information to localize the target speaker. While effective, these systems necessitate specific 
multi-microphone array configurations, limiting their versatility across diverse hardware.

Ultimately, regardless of the target speaker clues utilized, most existing frameworks overlook 
the practical challenge of obtaining high-quality references in spontaneous dialogue. 
The dependency on pre-collected data or specialized hardware remains a bottleneck 
for seamless human-machine interaction. To address these limitations, this paper 
investigates the feasibility of TSE using only the intrinsic information available during 
the interaction process itself. Specifically, we propose an EoW-TSE that directly adopts 
the wake-word segment as the enrollment reference. This approach eliminates the user's burden 
of manual enrollment.
The main contributions are summarized as follows: 
\begin{itemize}

\item We introduce the Enroll-on-Wakeup (EoW) TSE paradigm, utilizing the triggering wake-word as an automatic enrollment clue to facilitate zero-effort, seamless human-machine interaction.

\item We present the first systematic study of EoW-TSE, providing a comprehensive evaluation 
of advanced generative and discriminative models under diverse acoustic conditions and 
complex interferences;

\item We investigate enrollment augmentation methods using LLM-based TTS, demonstrating that synthetic enrollment significantly enhances perceptual quality under severe acoustic degradation, while identifying the remaining challenges in balancing speech intelligibility with ASR accuracy.

\end{itemize}

\vspace{-0.3cm}
\section{Problem Definition}
\label{sec:probdef}

\begin{figure*}
\centering
\includegraphics[width=1.0\linewidth]{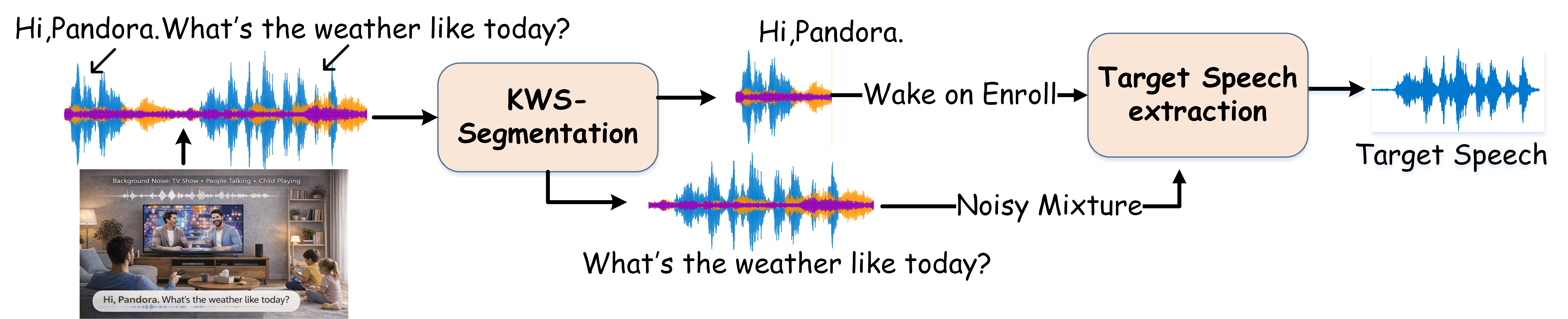}
    \caption{Illustration of EoW-TSE system.}
    \label{fig:eow}
    \vspace{-0.3cm}
\end{figure*}


This section defines the \textit{Enroll-on-Wakeup} (EoW) TSE framework, illustrated in Figure~\ref{fig:eow}. The goal of EoW-TSE is to extract target speech from a continuous noisy mixture by using only the transient wake-word segment as the enrollment reference.

\vspace{-0.2cm}
\subsection{Conventional TSE Formulation}

In a traditional TSE task, the observed noisy mixture $x(t)$ in the time domain is 
typically modeled as:
\begin{equation}
    x(t) = s(t) + n(t)
\end{equation}
where $s(t)$ represents the clean target speech and $n(t)$ denotes the sum of additive 
noise and interfering talkers. Standard models isolate $s(t)$ by leveraging a 
high-quality enrollment utterance $e_{pre}$, which is pre-recorded under controlled, 
clean conditions. The extraction process can be expressed as:
\begin{equation}
    \hat{s}(t) = \mathcal{F}(x(t) \mid e_{pre}; \Theta)
\end{equation}
where $\mathcal{F}$ is the TSE mapping function and $\Theta$ represents the model 
parameters. As discussed in Section~\ref{sec:intro}, the reliance on a 
pre-collected $e_{pre}$ limits the system's spontaneity and disrupts 
the user experience.

\vspace{-0.2cm}
\subsection{The EoW-TSE Architecture and Challenges}

Unlike the conventional approach, the proposed EoW-TSE derives the 
target speaker enrollment clue directly from the interaction process. 
As shown in Figure~\ref{fig:eow}, the input stream contains the wake-up command 
(e.g., ``Hi, Pandora'') followed immediately by the target query 
(e.g., ``What's the weather like today?''). 
The workflow of EoW-TSE can be defined as follows:

\begin{itemize}
    \item \textbf{KWS-Segmentation:} A keyword spotting (KWS) module segments the incoming stream into the wake-word segment $x_{wake}$ and the subsequent noisy mixture $x_{query}$;
    
    \item \textbf{Enroll-on-Wakeup:} The system automatically utilizes $x_{wake}$ as the EoW reference. Unlike the static $e_{pre}$, $x_{wake}$ is inherently characterized by limited duration and acoustic corruption with environmental noise and interference;
    
    \item \textbf{Target Extraction:} The TSE model uses this noisy, short-duration enrollment to extract the target speech $\hat{s}_{query}$ from the query mixture:
    \begin{equation}
        \hat{s}_{query} = \mathcal{F}(x_{query} \mid x_{wake}; \Theta)
    \end{equation}
\end{itemize}

The core innovation of EoW-TSE lies in its \textbf{zero-effort enrollment}, which 
enables seamless interaction without the conventional ``enroll-then-extract'' constraint. 
However, this architecture shift introduces significant challenges: the brevity of 
$x_{wake}$ falls to provide robust target speaker clues, and real-time environmental 
interference leads to ``clue contamination'', resulting in significant performance degradation.


\section{Enroll-on-Wakeup TSE Systems}
\label{sec:eowsystems}

This section details our experimental framework for 
evaluating EoW-TSE. We describe five real-world recording 
scenarios capturing diverse acoustic environments, followed 
by the four advanced TSE models used in our comparative 
study. Finally, we introduce the integration of LLM-based 
TTS engines for enrollment augmentation.

\vspace{-0.3cm}
\subsection{Scenario Description}
\label{subsec:scenarios}

\begin{table}[ht]
  \caption{Acoustic conditions of five EoW-TSE test scenarios. `$d$' denotes distance to the microphone (m), `$RT$' is the reverberation time ($RT_{60}$ in seconds), and `*-10/-5' indicates SNR levels of 10dB and 5dB respectively.}
  \label{tab:scenarios}
  \centering
  \resizebox{\columnwidth}{!}{
  \begin{tabular}{lllcc}
    \toprule
    \textbf{Scenario}      & \textbf{Condition}           & \textbf{\#spk/\#utt}    & \multicolumn{2}{c}{\textbf{Average duration (s)}}    \\
    & & & enrollment & mixture \\
    \midrule
    CloseNoise-10          & d=1, RT=0.4            & 45/503  & 1.08  & 1.78                        \\
    FarNoise-10            & d=3, RT=0.4            & 45/502  & 1.08  & 1.78                        \\
    FarNoiseReverb-10      & d=3, RT=0.6            & 30/278  & 1.17  & 1.79                     \\
    FarNoise-5             & d=3, RT=0.4            & 45/503  & 1.08  & 1.78                          \\
    FarNoiseReverb-5       & d=3, RT=0.6            & 15/226  & 0.97  & 1.76                        \\
    \bottomrule
  \end{tabular}}
  \vspace{-0.4cm}
\end{table}

To evaluate the robustness of EoW-TSE in real-world conditions, we 
utilize an internal dataset collected by Unisound\footnote{https://www.unisound.com/} 
consisting of real-world recordings across five distinct acoustic scenarios. 
These scenarios are designed to reflect the complexities of actual 
human-machine dialogue, varying in terms of speaker distance ($d$), 
reverberation time ($RT_{60}$), and signal-to-noise ratio (SNR). 
The detailed statistics are summarized in 
Table~\ref{tab:scenarios}. The specific configurations and challenges are as follows:

\begin{itemize}
    \item \textbf{Interference and Noise Types:} 
    In the \textit{CloseNoise} scenario, the recording is conducted in a relatively clean environment. 
    In contrast, the other four scenarios (\textit{FarNoise} and \textit{FarNoiseReverb} variants) 
    feature a challenging acoustic background consisting of audio from a television variety show. Both 
    the enrollment segments ($x_{wake}$) and the noisy mixtures ($x_{query}$) in these scenarios contain 
    complex TV background noise, non-target human speech from the program, and additional spontaneous 
    interfering talkers in the recording room.
    
    \item \textbf{Acoustic Variabilities:} 
    We evaluate the impact of distance by comparing close-talk ($d=1$m, \textit{CloseNoise}) and 
    far-field ($d=3$m, \textit{FarNoise}) configurations. We also assess the 
    impact of increased reverberation by varying $RT_{60}$ from 0.4s to 0.6s (\textit{FarNoiseReverb}). 
    To further test the noise robustness of TSE models, we include both 10dB and 5dB SNR levels 
    under these far-field conditions.
    
    \item \textbf{Wake-up Phrases:} The enrollment utilize two specific Chinese wake-up commands: 
    ``Hi, Pandora'' or ``Hello, Cube''. 
    
\end{itemize}

As detailed in Table~\ref{tab:scenarios}, our test set includes a diverse 
collection of speakers, with 15 to 45 unique identities per scenario and over 
2,000 utterances in total. Notably, the average enrollment duration ($x_{wake}$)
is approximately 1.0s. This is significantly shorter than the 
multi-second or multi-utterance enrollments used in conventional TSE benchmarks. 
This brief duration, combined with the aforementioned real-world interferences, 
highlights the critical target speaker guidance information scarcity and 
contamination challenges of the EoW-TSE.

\subsection{Target Speech Extraction Systems}
\label{subsec:tse}


In this study, we evaluate four advance TSE models proposed in recent years 
to assess their performance under the EoW-TSE paradigm, including three 
discriminative systems (SEF-PNet\cite{huang2025sef}, LExt\cite{Shen2025ListenTE}, and CIE-mDPTNet \cite{yang2024target}) and one generative framework (SoloSpeech
\cite{wang2025solospeech}). 

\subsubsection{Discriminative TSE Models}
\label{subsubsec:dtse}

\textbf{SEF-PNet}\cite{huang2025sef} is a speaker-embedding-free TSE model that directly 
integrates enrollment and mixture using an interactive speaker adaptation (ISA) 
module plus a local–global context aggregation (LCA) mechanism. 
By performing iterative T-F domain interaction between the enrollment and mixture, 
the model achieves robust personalized speech enhancement without relying on external 
speaker encoders or pre-trained speaker models. It has demonstrated superior performance 
across all three TSE conditions on the Libri2Mix\cite{cosentino2020librimix} benchmark.

\textbf{LExt}\cite{Shen2025ListenTE} simply prepends an enrollment utterance (plus a tiny glue segment) 
to the mixture waveform so the network “hears” the target speaker first. 
The model learns to use that artificial onset as a prompt to extract the later target speech. 
This remarkably simple training-time concatenation trick works with standard 
separator architectures (such as TF-GridNet\cite{wang2023tf} or TF-LocoFormer\cite{saijo2024tf}) 
and achieves strong TSE performance on WSJ0-2mix\cite{hershey2016deep} and WHAM!\cite{wichern2019wham}/ WHAMR!\cite{maciejewski2020whamr}.

\textbf{CIE-mDPTNet}\cite{yang2024target} focuses on directly exploiting contextual information 
in the T-F domain to overcome the limitations of insufficient context utilization. 
It employs a simple attention mechanism to compute correlation weights between the 
enrollment speech and the mixture. Combined with a dual-path transformer architecture\cite{chen2020dual}, 
it effectively captures both short-term spectral variations and long-term temporal dependencies, 
even in complex, multi-speaker TSE environments.

\subsubsection{Generative TSE Model}

\textbf{SoloSpeech}\cite{wang2025solospeech} is a cascaded generative TSE pipeline 
comprising an audio compressor, a speaker-embedding-free extractor, and a T-F domain 
diffusion corrector. By operating in the latent space and employing an iterative 
correction process, it effectively overcomes the artifacts and naturalness 
degradation common in discriminative methods. It achieves SOTA 
results in both intelligibility and perceptual quality, demonstrating exceptional 
robustness in out-of-domain and real-world scenarios.

\subsection{Augment Enrollment  with LLM-based TTS Models}
\label{subsec:enrollaug}

The core challenge of EoW-TSE lies in the target guidance information 
scarcity and contamination, as enrollment signals consist of 
extremely short and noisy wake-up fragments. To address this, 
three SOTA zero-shot generative TTS models are employed: the \textbf{IndexTTS2} \footnote{\href{https://huggingface.co/IndexTeam/IndexTTS-2}{ https://huggingface.co/IndexTeam/IndexTTS-2}} 
\cite{zhou2025indextts2}, \textbf{xTTS} \footnote{\href{https://huggingface.co/coqui/XTTS-v2/tree/v2.0.2}{https://huggingface.co/coqui/XTTS-v2/tree/v2.0.2}} \cite{casanova2024xtts}, and 
\textbf{CosyVoice3} \footnote{\href{https://huggingface.co/FunAudioLLM/Fun-CosyVoice3-0.5B-2512}{https://huggingface.co/FunAudioLLM/Fun-CosyVoice3-0.5B-2512}}\cite{du2025cosyvoice}. 
By using $x_{wake}$ as an acoustic prompt, these models $G(\cdot)$ perform 
target-identity-preserving synthesis to generate high-fidelity, clean enrollment speech. 
Two enrollment augmentation methods are explored:
1) \textbf{Clean Re-synthesis (CR):} We re-synthesize the original wake-up transcript $t_{wake}$ to produce a clean version $\hat{x}_{wake} = G(t_{wake} | x_{wake})$, which serves as the sole enrollment for EoW-TSE extraction;  
2) \textbf{Extended Concatenation (EC):} To increase identity-clue diversity, we generate an auxiliary clean segment $a_{gen} = G(t_{gen} | x_{wake})$ using a ChatGPT-generated sentence $t_{gen}$. This is then concatenated with the original segment to form an extended enrollment: $e_{aug} = [y_{wake} \oplus a_{gen}]$. 

These methods aim to evaluate whether substituting or augmenting contaminated enrollment fragments 
with synthesized clean speech can effectively stabilize the target speaker guidance 
under the EoW-TSE scenario.

\vspace{-0.1cm}
\section{Experimental Setup}
\label{sec:expset}

\textbf{Datasets}: 
We evaluate all models on the five-scenario test set detailed in 
Table~\ref{tab:scenarios} at a 16 kHz sampling rate. For the discriminative 
systems: SEF-PNet, CIE-mDPTNet, and LExt, they all are trained on the 
Libri2Mix \textit{train-100} subset under 
the \textit{mix\_both} (2-speaker + noise) and \textit{min} duration modes. 
For the generative SoloSpeech, we used the original authors' pre-trained 
checkpoint\footnote{\href{https://wanghelin1997.github.io/SoloSpeech-Demo/}{https://wanghelin1997.github.io/SoloSpeech-Demo/}}, 
which was trained on the Libri2Mix \textit{train-360} subset under 
identical mixture conditions.

\textbf{Configurations:} 
SEF-PNet and CIE-mDPTNet were trained for 
130 epochs using the Adam\cite{kinga2015method} optimizer with an initial learning rate (LR) of 5e-4 and 
L2-norm gradient clipping\cite{zhang2019gradient}. The LR was decayed by 0.98 every two epochs for 
the first 100 epochs, and by 0.9 thereafter. LExt used an initial LR of 1e-4 with a 
weight decay of 1e-5. For its TF-GridNet backbone, we adopted a more compact 
configuration than the original TF-GridNetV1 \cite{Shen2025ListenTE} to 
ensure a fast training.

\textbf{Evaluation Metrics:}
We employ SI-SDR\cite{le2019sdr}, PESQ\cite{rix2001perceptual}, and STOI\cite{taal2010short} to evaluate signal fidelity and quality, while DNSMOS\cite{reddy2022dnsmos} and word error rate (WER) assess perceptual performance and intelligibility across the five real-world scenarios. All WER results are calculated via the Fun-ASR\footnote{\href{https://huggingface.co/FunAudioLLM/Fun-ASR-Nano-2512}{https://huggingface.co/FunAudioLLM/Fun-ASR-Nano-2512}}\cite{an2025fun} service and the \texttt{meeteval} toolkit \cite{vonneumann2024meetevaltoolkitcomputationword} to ensure consistent scoring across diverse acoustic and segmentation conditions.

\begin{table}[!htbp]
  \caption{Results on Libri2Mix 2-Speaker+Noise. $\star$ are the results reproduced by us.}
  \label{tab:librimix}
  \centering
    \resizebox{\columnwidth}{!}{
    \begin{tabular}{lcccccc}
    \toprule
    \textbf{Models} & \textbf{SI-SDR} & \textbf{PESQ} & \textbf{STOI} & \textbf{Params} & \textbf{MACs} \\
    \midrule
    Mixture & -1.96 & 1.08 & 64.73 & - & - \\
    \midrule
    SEF-PNet \cite{huang2025sef} & 8.18 & 1.55 & 82.67 & 6.08M & \textbf{15.87G} \\
    CIE-mDPTNet \cite{yang2024target} & 9.47 & 1.78 & 85.35 & \textbf{2.9M} & 48.3G \\
    $\star$LExt \cite{Shen2025ListenTE} & 10.47 & 1.88 & \textbf{87.26} & 3.9M & 61.0G \\
    SoloSpeech \cite{wang2025solospeech} & \textbf{11.12} & \textbf{1.89} & - & 590.9M & - \\
    \bottomrule
    \end{tabular}}
    \vspace{5pt}
    \begin{minipage}{0.98\columnwidth}
    \footnotesize MACs are theoretically computed based on model architecture.
  \end{minipage}
    \vspace{-0.3cm}
\end{table}

\vspace{-0.2cm}
\begin{table*}[!htbp]
    \centering
    \caption{Results on enrollments synthesized using different TTS models.}
    \label{tab:enroll-synthetic}
    \scalebox{0.8}{
    \begin{tabular}{l|cccc|cccc} 
        \toprule
        \multicolumn{1}{c}{\textbf{Conditions}} & \multicolumn{4}{c}{\textbf{DNSMOS} $\uparrow$} & \multicolumn{4}{c}{\textbf{WER} $\downarrow$} \\
         & Noisy & xTTS & IndexTTS2 & CosyVoice3 & Noisy & xTTS & IndexTTS2 & CosyVoice3   \\
        \midrule
        CloseNoise-10  & 2.029 & 2.472 & 2.564 & \textbf{2.609} & 14.61 & 39.76 & \textbf{11.23} & 40.36 \\
        FarNoise-10 & 1.590 & \textbf{2.464} & 2.237 & 2.428 & 27.74 & 40.04 & \textbf{19.27} & 54.88\\
        FarNoiseReverb-10 & 1.262 & \textbf{2.549} & 1.917 & 2.114 & 19.69 & 33.45 & \textbf{16.10} & 59.71 \\
        FarNoise-5 & 1.440 & 2.274 & 2.063 & \textbf{2.400} & 45.13 & 46.57 & \textbf{36.58} & 80.62 \\
        FarNoiseReverb-5 & 1.251 & \textbf{2.473} & 1.743 & 2.107 & 70.58 & 55.86 & \textbf{44.47} & 102.54 \\
        \bottomrule
    \end{tabular}}
    \vspace{-0.2cm}
\end{table*}

\begin{table*}[!htbp]
    \centering
    \caption{EoW-TSE results with different enrollment augmentation methods on CIE-mDPTNet. }
    \label{tab:enroll-cie}
       \scalebox{0.8}{
    {\begin{tabular}{l|ccccc|ccccc}
        \toprule
        \multicolumn{1}{c}{\textbf{Conditions}} & \multicolumn{5}{c}{\textbf{DNSMOS} $\uparrow$} & \multicolumn{5}{c}{\textbf{WER} $\downarrow$} \\
         & Noisy & xTTS (CR) & IndexTTS2 (CR) & xTTS(EC) & IndexTTS2(EC) & Noisy & xTTS(CR) & IndexTTS2(CR) & xTTS(EC) & IndexTTS2(EC)  \\
        \midrule
        CloseNoise-10  & 1.904 & 2.163 & 2.119 & \textbf{2.197} & 2.170 & \textbf{3.10} & 4.37 & 3.24 & 3.91 & 3.48 \\
        FarNoise-10 & 1.371 & 1.511 & 1.481 & \textbf{1.527} & 1.500 & \textbf{4.54} & 6.41 & 5.38 & 6.81 & 5.14 \\
        FarNoiseReverb-10 & 1.165 & 1.202 & 1.200 & \textbf{1.214} & 1.195 & \textbf{3.52} & 7.15 & 5.06 & 7.00 & 6.08 \\
        FarNoise-5 & 1.286 & 1.379 & 1.366 & \textbf{1.396} & 1.371 & \textbf{13.31} & 16.01 & 29.55 & 15.92 & 25.61 \\
        FarNoiseReverb-5 & 1.173 & 1.216 & \textbf{1.223} & 1.217 & 1.212 & \textbf{15.95} & 22.89 & 31.32 & 23.12 & 28.36 \\
        \bottomrule
    \end{tabular}}}
    \vspace{-0.4cm}
\end{table*}

\vspace{-0.1cm}
\section{Results and Discussion}
\label{sec:rst}

\subsection{Results on Libri2mix 2spk+noise Condition}
\label{subsec:baserst}

Table~\ref{tab:librimix} shows the comparative performance of the selected models 
on the challenging Libri2Mix 2-speaker+noise condition. We evaluate a diverse 
set of architectures varying in parameter size and computational complexity (MACs). 
Among discriminative models, LExt achieves the best performance. 
The generative SoloSpeech attains SOTA results in both SI-SDR and PESQ. 
Overall, all selected models exhibit competitive performance, establishing a robust 
baseline for investigating their behavior under the more constrained EoW-TSE scenarios.

\subsection{Overall Performance of EoW-TSE Systems}
\label{subsec:overall}

\begin{figure}[h]
  \centering
  \includegraphics[width=\linewidth]{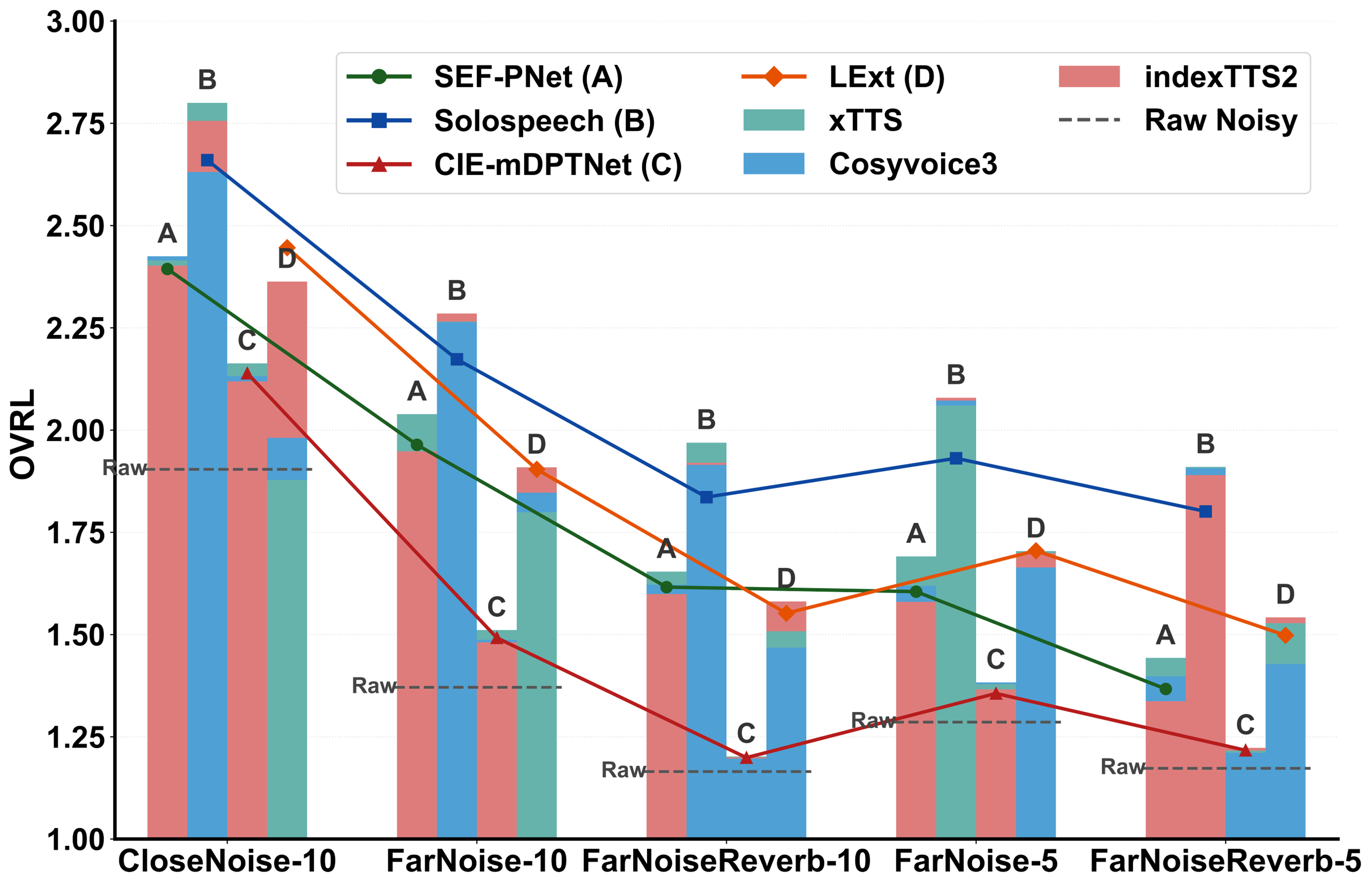}
  \caption{OVRL scores on five scenarios: Original EoW-TSE vs. TTS-augmented (CR).}
  \label{fig:test_ovrl}
  \vspace{-0.7cm}
\end{figure}

\begin{figure}[h]
  \centering
  \includegraphics[width=\linewidth]{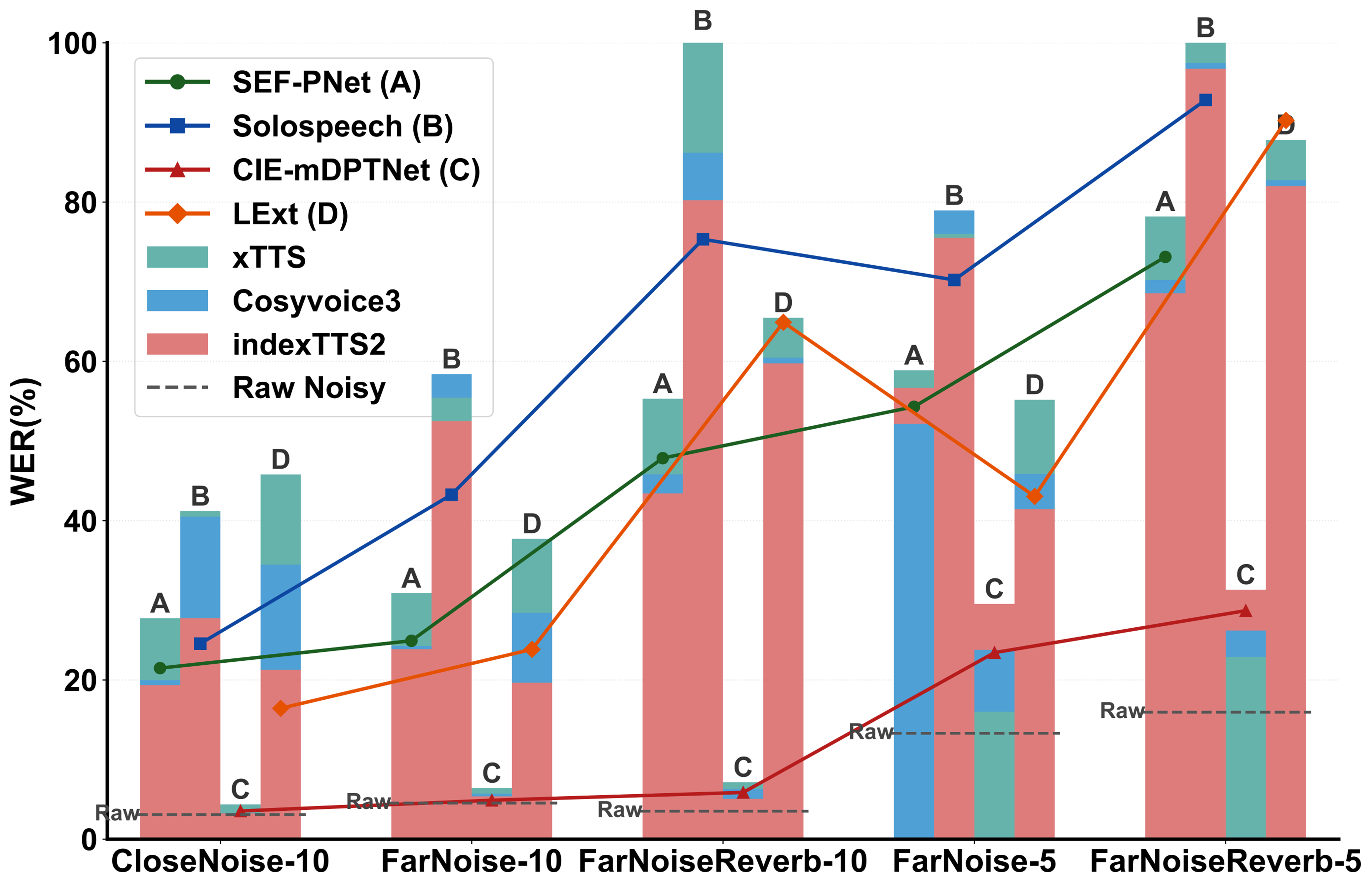}
  \caption{WERs on five scenarios: Original EoW-TSE vs. TTS-augmented (CR).}
  \label{fig:test_wer}
  \vspace{-0.7cm}
\end{figure}

Figures~\ref{fig:test_ovrl} and \ref{fig:test_wer} show the results of four advanced 
TSE models under the EoW-TSE paradigm. Regarding perceptual quality, the generative SoloSpeech 
consistently achieves the highest OVRL scores across all scenarios, 
followed by SEF-PNet and LExt with comparable performance, while CIE-mDPTNet lags behind. 
However, this trend reverses in the speech intelligibility measured by WER. 
CIE-mDPTNet demonstrates the most robust ASR performance, whereas SoloSpeech 
suffers from a dramatic WER increase as environmental complexity intensifies 
(e.g., in far-field and reverberant conditions). Notably, none of these TSE models 
outperforms the raw noisy mixture's direct WER in these challenging scenarios. 
This discrepancy suggests that while generative models are great at generating high-quality speech, they may introduce phonetic 
distortions that hinder ASR accuracy, highlighting a critical area 
for future optimization in EoW-TSE research.

\subsection{Effect of Synthetic Enrollment}
\label{subsec:synthes}

Results in Table~\ref{tab:enroll-synthetic} aim to assess the quality of 
synthetic enrollment generated by IndexTTS2, xTTS and CosyVoice3. 
It is clear to see that while xTTS and CosyVoice3 achieve superior DNSMOS, 
only IndexTTS2 consistently reduces the WER compared to raw noisy 
enrollment. This trend is mirrored in Figures~\ref{fig:test_ovrl} and \ref{fig:test_wer},
where IndexTTS2-augmented models (red bars) significantly outperform other synthetic variants in 
speech intelligibility, effectively stabilizing the TSE output.

Table~\ref{tab:enroll-cie} further explores the impact of the proposed enrollment 
augmentation methods on the most WER robust TSE system, CIE-mDPTNet. 
While both Clean Re-synthesis (CR) and Extended Concatenation (EC) consistently improve the \textit{DNSMOS} of the extracted target speech compared to the noisy mixture, 
neither method succeeds in reducing the \textit{WER}. This trend aligns with the findings in Figure~\ref{fig:test_wer}, confirming that existing TSE models introduce significant phonetic distortion while suppressing noise, thereby hindering ASR performance even with enhanced guidance. Notably, although the EC method marginally improves perceptual quality over CR, it offers comparable \textit{WER}, suggesting that simply increasing enrollment duration does not effectively compensate for these intelligibility losses. These results underscore a critical challenge in EoW-TSE: while zero-shot LLM-based TTS effectively mitigates guidance contamination, balancing perceptual enhancement with linguistic fidelity remains an open problem.

\vspace{-0.1cm}
\section{Conclusion}
\label{sec:con}

This paper presents the first systematic study of TSE under the Enroll-on-Wakeup paradigm. 
Our evaluation across five real-world scenarios reveals a clear perceptual-recognition gap: while 
advanced TSE models, especially generative ones, can notably improve audio quality, they often fail 
in ASR performance. We show that current TSE systems fall short of ideal performance when limited to 
a one-second wake-up reference. Furthermore, while TTS helps reduce enrollment noise, 
the trade-off between speech perceptual and intelligibility remains a major bottleneck. 
These findings establish a key baseline and offer practical guidelines for building next-generation robust EoW-TSE systems for seamless human-machine interaction.

\section{Generative AI Use Disclosure}
\label{sec:genAI}

The authors only used an LLM-based tool (ChatGPT) to generate sentence texts randomly as input for enrollment augment with Extended Concatenation (EC) models. The ideas, experimental design, code, analysis, and results of this study did not use any generative artificial intelligence techniques, and all content is original to the authors.

\bibliographystyle{IEEEtran}
\bibliography{main}

\end{document}